\documentclass[12pt]{iopart}
\usepackage{graphicx}
\usepackage{dcolumn}
\usepackage{bm}
\usepackage{epsf}

\begin{document}

\title{Level Splitting in Association with the Multiphoton Bloch-Siegert
  Shift}
\author{P L Hagelstein$^1$, I U Chaudhary$^2$}

\address{$^1$ Research Laboratory of Electronics, 
Massachusetts Institute of Technology, 
Cambridge, MA 02139,USA}
\ead{plh@mit.edu}

\address{$^2$ Research Laboratory of Electronics, 
Massachusetts Institute of Technology, 
Cambridge, MA 02139,USA}
\ead{irfanc@mit.edu}

\begin{abstract}
We present a unitary equivalent spin-boson Hamiltonian in which terms can be
identified which contribute to the Bloch-Siegert shift, and to the level
splittings at the anticrossings associated with the Bloch-Siegert resonances.
First-order degenerate perturbation theory is used to develop approximate
results in the case of moderate coupling for the level splitting.
\end{abstract}

\submitto{\jpb}

\maketitle

\section{Introduction}
\label{sec:intro}

The dynamics of a two-level with sinusoidal coupling has been of interest since the
time of Bloch and Siegert \cite{BlochSiegert,Shirley}.  
The (closely related) basic model for a two-level system coupled to a simple harmonic oscillator was considered 
 by Cohen-Tannoudji et al \cite{Cohen}.
The coupling in these models produces an increase in the two-level system transition energy 
(sometimes termed the Bloch-Siegert shift).
As the coupling strength is increased, the levels shift relative to one another, producing
both level crossings and level anticrossings.
Level crossings occur when the dressed two-level transition energy matches an even number of
oscillator quanta (in which case the parity of the states are mismatched, so no mixing occurs). 
Level anticrossings occur when the dressed two-level transition energy 
is resonant with an odd number of oscillator quanta, with the magnitude
of the splitting indicative of the ability of the coupled system to
convert energy between the two different degrees of freedom.

These models were studied initially in the context of spin dynamics in a magnetic field \cite{BlochSiegert,Shirley},
but they also appear in other applications.  
The coupling between atoms and an electromagnetic field can in some cases be described by these models,
in which case the resonances mentioned above correspond to multiphoton interactions.
Multiphoton resonances in which a substantial number of photons are exchanged have become
experimentally accessible recently \cite{Fregenal}. 
In part because of this there has been renewed interest in the multiphoton regime
\cite{Forre,OstrovskyHorsdal}.

We have found a unitary transformation which produces a rotated version of the problem
which appears to provide a clean separation between terms which produce most of the Bloch-Siegert
shift, and terms which produce the level splitting at the anticrossings.
This is interesting because it allows us to develop estimates when the coupling is moderately strong
for both the shift and the splittings using conventional methods on the rotated problem.
In essence, we are able to capture most of the level splitting in the multiphoton regime in terms
of first-order coupling in the context of degenerate perturbation theory.
This provides a new way to look at the problem which may be useful.

\section{Unitary equivalent Hamiltonian}

The Hamiltonian for the coupled two-level system and oscillator of interest
(the spin-boson Hamiltonian) can be written as

\begin{equation}
\label{eq:BasicHamiltonian}
\hat{H} 
~=~
{ \Delta E  \over 2} \hat{\sigma}_z
+
\hbar \omega_0 \hat{a}^\dagger \hat{a}
+
U (\hat{a}^\dagger+\hat{a}) \hat{\sigma}_x
\end{equation}

\noindent 
where the $\hat{\sigma}_i$ are the Pauli matrices. 
Since we are interested in the multiphoton regime, we assume that
the background excitation of the photon field is large:  

\[
\Delta E \gg \hbar \omega_0 \; , \; \; \; n \gg 1
\]

Rotations are often used to simplify Hamiltonians \cite{Wagner}; however, in
this case our rotation will make the problem more complicated mathematically
but perhaps simpler functionally as outlined above.
We consider the unitary equivalent Hamiltonian

\begin{equation}
\hat{H}^\prime 
~=~ \hat{\mathcal{U}}^\dagger \hat{H} \hat{\mathcal{U}}
\end{equation}

\noindent
where 

\[
\hat{\mathcal{U}} 
~=~ 
\exp \left \lbrace - \frac{i}{2} 
          \arctan \left \lbrack
       {2 U (\hat{a} + \hat{a}^\dagger) \over \Delta E} 
                  \right \rbrack
       \hat{\sigma}_y \right \rbrace 
\]

\noindent 
The rotated Hamiltonian, $\hat{H}^\prime$ can be broken up into an unperturbed
part ($\hat{H}_0$), and pieces which will be considered as perturbations  ($\hat{V}$ and $\hat{W}$):

\begin{equation}
\hat{H}^\prime 
~=~
\hat{H}_0
~+~
\hat{V}
~+~
\hat{W}
\end{equation}

\noindent
where 

{\small

\begin{equation}
\hat{H}_0
~=~
\sqrt{\Delta E^2 + 4 U^2 (\hat{a}+\hat{a}^\dagger)^2 } {\hat{\sigma}_z
  \over 2}
~+~
\hbar \omega_0 \hat{a}^\dagger \hat{a} 
\end{equation}

\[
\hat{V} = 
{i \hbar \omega_0 \over 2}
\left \lbrace
\left [
{ \displaystyle{  U  \over \Delta E}  \over 1+  \left [ \displaystyle{ 2U 
(\hat{a}+\hat{a}^\dagger) \over \Delta E} \right ]^2  }
\right ]
(\hat{a}-\hat{a}^\dagger)
\right .
\]
\begin{equation}
\ \ \ \ \ \ \ \ \ \ \ \ \ \ \ \ \ \ \ \ \ \ \ \ \ \ \ \ \ \ \ \ \ \ \ \
\left .
~+~
(\hat{a}-\hat{a}^\dagger)
\left [
{ \displaystyle{  U  \over \Delta E}  \over 1+   \left [ \displaystyle{ 2 U 
(\hat{a}+\hat{a}^\dagger) \over \Delta E} \right ]^2  }
\right ]
\right \rbrace
\hat{\sigma}_y 
\end{equation}

\begin{equation}
\hat{W} = \hbar \omega_0 
\left \lbrace
{ \displaystyle{  U  \over \Delta E}  \over 1+   \left [ \displaystyle{
      2 U (\hat{a}+\hat{a}^\dagger) \over \Delta E} \right ]^2  } 
\right \rbrace^2
\end{equation}

}

\noindent
In the multiphoton regime of interest here ($n \gg 1$ and $\Delta E \gg \hbar \omega_0$),
the last term produced by the rotation, $\hat{W}$, is small; we will therefore neglect
it in what follows.

\section{Eigenvalues of $\hat{H}_0$}

Consider first the eigenvalue equation of the unperturbed Hamiltonian 
$\hat{H}_0$ in
the rotated frame

\begin{equation}
E \psi ~=~ \hat{H}_0 \psi
\end{equation}

\noindent
Separation of variables allows us to develop solutions of the form

\begin{equation}
\psi ~=~ u \otimes |s,m \rangle
\end{equation}

\noindent
where $u$ satisfies

{\small

\begin{equation}
\left ( E + {\hbar \omega_0 \over 2} \right ) u(y)
~=~ 
{\hbar \omega_0 \over 2} \left [-{d^2 \over dy^2} + y^2 \right ]u(y)
+
m \sqrt{\Delta E^2 + 8 V^2 y^2 } u(y)
\end{equation}

}

\noindent
Both from numerical calculations and the WKB approximation we have found
that the energy eigenvalues are given approximately by

\begin{equation}
E_{n,m}(g) ~=~ \Delta E(g)m + \hbar \omega_0n
\end{equation}

\noindent
The WKB approximation can be used to develop a useful analytic approximation to the
dressed two-level transition energy $\Delta E(g)$, which we may write as 

\begin{equation}
\Delta E(g)
~=~
{\Delta E \over \pi}
\int_{-\sqrt{\epsilon}}^{\sqrt{\epsilon}}
\sqrt{
1 + 8 g^2 y^2/n 
\over 
\epsilon - y^2
}
dy
\end{equation}

\noindent
with $\epsilon = 2n+1$.  The dimensionless coupling constant $g$ is

\begin{equation}
g ~=~ {U \sqrt{n} \over \Delta E}
\end{equation}

\noindent
In the limit of large $n$ and $\hbar \omega_0 \ll \Delta E$ where this is valid, 
the rotated system governed by $\hat{H}_0$ alone behaves like a dressed two-level 
system (with increased transition energy) and an unperturbed oscillator.

The condition for Bloch-Siegert resonances can be written as 

\begin{equation}
\Delta E(g) ~=~ (2k + 1)  \hbar \omega_0
\end{equation}

\noindent
Level crossings occur in the modified version of the problem described
by the unperturbed Hamiltonian $\hat{H}_0$ in the rotated frame.  As level anticrossing
occur in the original spin-boson model at these resonances, the coupling that is responsible
for the level anticrossing has been eliminated in $\hat{H}_0$.  This is an interesting
and perhaps unexpected feature of this rotation.

\section{Level splitting in the unrotated Hamiltonian}

Near a resonance, we can use a two-level description to account for the
level splittings.

\begin{equation}
E(g)
\left (
\begin{array} {c}
c_0   \cr
c_1 \cr
\end{array}
\right )
~=~
\left (
\begin{array} {cc}
E_0(g) & v   \cr
v   & E_1(g) \cr
\end{array}
\right )
\left (
\begin{array} {c}
c_0   \cr
c_1 \cr
\end{array}
\right )
\end{equation}

\noindent
Two levels with energies $E_0$ and $E_1$ that depend on the
dimensionless coupling strength $g$ cross, and couple to each other 
with an interaction $v$ which we assume to be constant in the vicinity 
of the resonance.  At resonance ($g_0$), the two levels 
in this simplified model are degenerate

\begin{equation}
E_0(g_0) ~=~ E_1(g_0)
\end{equation}

\noindent
The splitting between the two levels at this point is twice the
magnitude of the interaction 

\begin{equation}
\Delta E_{min} ~=~ E_+(g_0) - E_-(g_0) ~=~ 2 |v|
\end{equation}

The level splittings in the case of weak coupling have been known for some 
time \cite{Shirley}, as mentioned above.  Shirley's results written in our
notation are   

\begin{equation}
\Delta E_{min} ~=~ \frac{g_0^{2k+1}}{2^{2k-1} (k!)^2} \left(\frac{\Delta E}{\hbar
  \omega_0}\right)^{2k} \Delta E
\end{equation}

\begin{figure} [h]
\begin{center}
\mbox{\epsfbox{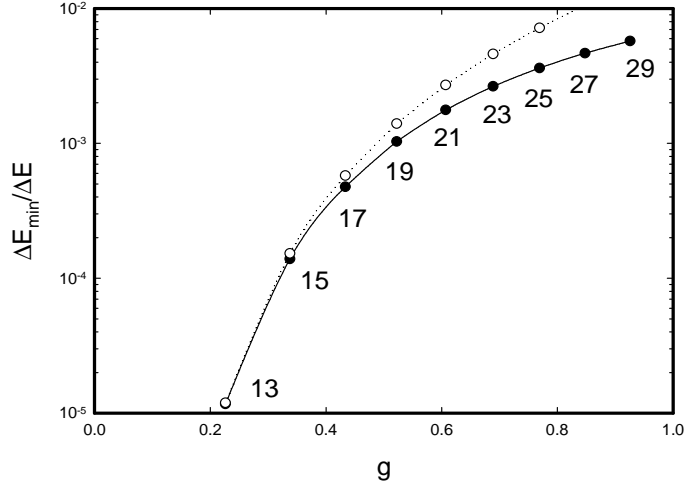}}
\caption{Energy level splitting on resonance as a function of the
  dimensionless coupling 
strength for $\Delta E = 11 \hbar \omega_0$ at large $n$.  Exact
  numerical results: full circles; literature results: open
  circles. The odd integers label the Bloch-Siegert resonance $2 k + 1$.} 
\label{Fig32x2a}
\end{center}
\end{figure}

\noindent
We have plotted results from the direct numerical solution of the spin-boson Hamiltonian
[Equation (\ref{eq:BasicHamiltonian})], and also for the this weak coupling result
in \fref{Fig32x2a}.  When the dimensionless coupling constant $g$ is small 
the results match well; 
when the coupling gets stronger, we see (as expected) that perturbation theory begins to 
break down.

\epsfxsize = 4.00in
\epsfysize = 2.70in

\section{Level splitting in the rotated Hamiltonian}

The dressed transition energy of the two-level system is described reasonably
well through the unperturbed part $\hat{H}_0$ of the rotated Hamiltonian, but
no level splittings occur in the eigenvalues of $\hat{H}_0$.
Hence, all of the splitting must be due to the terms we have considered to
be perturbations.
In this section, our goal is to apply degenerate perturbation theory in the
rotated frame to see whether the larger of the perturbation terms $\hat{V}$
can account for the level splitting.

To calculate the level splitting in the vicinity of an anticrossing, we need to 
compute the eigenkets $\psi_{n,m}$ of $\hat{H}_0$ where

\begin{equation}
\hat{H}_0 \psi_{n,m} ~=~ E_{n,m} \psi_{n,m}
\end{equation}

\noindent
This can be done numerically, or by using the WKB approximation (which we have found
to be effective for such problems).  Near the $(2 k + 1)$th resonance, the level anticrossing is
well-described by a simple two-level approximation

{\small

\begin{equation}
E(g)
\left (
\begin{array} {c}
c_0   \cr
c_1 \cr
\end{array}
\right )
~=~
\left (
\begin{array} {cc}
E_{n,m}(g) & \langle \psi_{n,m} | \hat{V} | \psi_{n+2k+1,m-1} \rangle   \cr
\langle \psi_{n+2k+1,m-1}|\hat{V}|\psi_{n,m} \rangle   & E_{n+2k+1,m-1}(g) \cr
\end{array}
\right )
\left (
\begin{array} {c}
c_0   \cr
c_1 \cr
\end{array}
\right )
\end{equation}

}

\noindent
The energy splitting at resonance is

\begin{equation}
\Delta E_{min} ~=~
E_+(g_0) - E_-(g_0) ~=~ 2|\langle \psi_{n,m} | \hat{V} | \psi_{n+2k+1,m-1} 
\rangle| 
\label{split}
\end{equation}

\epsfxsize = 4.00in
\epsfysize = 2.70in
\begin{figure} [t]
\begin{center}
\mbox{\epsfbox{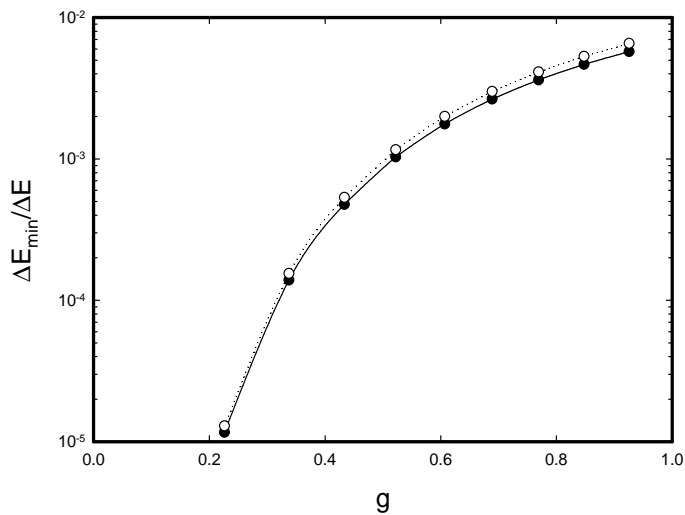}}
\caption{Energy level splitting on resonance as a function of the
  dimensionless coupling strength for $\Delta E = 11 \hbar \omega_0$ at large $n$.  
  Exact numerical results: full circles; first-order splitting from \eref{split}: open circles.} 
\label{Fig32x2}
\end{center}
\end{figure}

\noindent
In \fref{Fig32x2} we have plotted level splittings taken from a direct
numerical solution of the original spin-boson Hamiltonian [Equation (\ref{eq:BasicHamiltonian})]
and also from first-order degenerated perturbation theory as discussed here
(we used numerical solutions for the eigenfunctions $\psi_{n,m}$ for this result).
We can see from \fref{Fig32x2} that the exact numerical results
for the level splitting of the unrotated Hamiltonian match very well the
results obtained by using degenerate perturbation theory on the rotated
Hamiltonian. 
Minor deviations occur at the larger $g$ values which we attribute to the omission of
higher-order terms in the degenerate perturbation theory.

\section{Conclusion}

We have found a useful unitary transformation that produces a
rotated Hamiltonian for the spin-boson problem in the multiphoton 
regime that has interesting properties.  
The rotated Hamiltonian is more complicated mathematically than the 
initial spin-boson Hamiltonian, but appears to be simpler in terms of 
functionality.
One part of the rotated Hamiltonian is identified as an unperturbed
Hamiltonian ($\hat{H}_0$) which appears to describe the coupled systems
reasonably well away from the level anticrossings. 
This part of the problem is useful for developing estimates of the Bloch-Siegert shift.
Another part of the rotated Hamiltonian ($\hat{V}$) is identified as a perturbation
which is responsible for most of the coupling which occurs at the anti-crossing.
Used with first-order degenerate perturbation theory, this term provides a reasonable
approximation for the level splittings at the Bloch-Siegert resonances.
Finally, there is present an additional term ($\hat{W}$) in the rotated Hamiltonian 
which is small (so that we have neglected it in our discussion here), but which
can provide a minor correction to the dressed two-level system energies.


\section*{References}

\begin{thebibliography}{7}  
\bibitem{BlochSiegert} Bloch F and  Siegert A 1940 \PR {\bf 57} 522
\bibitem{Shirley} Shirley J 1965 \PR {\bf 138}, B979   
\bibitem{Cohen} Cohen-Tannoudji C, Dupont-Roc J, and Fabre C 1973 \JPB {\bf 6} L214
\bibitem{Fregenal}  Fregenal D \etal 2004 {\it Phys. Rev.} A {\bf 69}
  031401(R)
\bibitem{Forre}  F\o rre M 2004 {\it Phys. Rev.} A {\bf 70} 013406
\bibitem{OstrovskyHorsdal} Ostrovsky V N and Horsdal-Pedersen E 2004 {\it
  Phys. Rev. A} {\bf 70} 033413
\bibitem{Wagner} Wagner M 1986, {\it Unitary transformations in solid state
  physics} (New York: North-Holland)
\bibitem{AhmadBullough} Ahmad F and Bullough R K 1974 \JPB {\bf 7} L275  
\end{thebibliography}
\end{document}